\documentclass[5p,twocolumn,times,number]{elsarticle}

\usepackage{graphicx}
\usepackage{amsmath}
\usepackage{lineno}
\RequirePackage{xspace}
\usepackage{relsize}

\def\KS{\ensuremath{K^0_{\scriptscriptstyle S}}\xspace} 
\def\YnS{\ensuremath{\Upsilon{(nS)}}\xspace}
\def\etal  {{\it et~al.}}

\begin{document}

\begin{frontmatter}

\title{Belle II Silicon Vertex Detector}

\cortext[cor]{Corresponding author. Tel.: +91 22 22782147; Fax: +91 22 22804610}

\author[add6]{K.~Adamczyk}
\author[add17]{H.~Aihara}
\author[add11,add12]{C.~Angelini}
\author[add15]{T.~Aziz}
\author[add15]{V.~Babu}
\author[add6]{S.~Bacher}
\author[add1]{S.~Bahinipati}
\author[add8]{E.~Barberio}
\author[add8]{Ti.~Baroncelli}
\author[add8]{To.~Baroncelli}
% ...
\address[add1]{Indian Institute of Technology Bhubaneswar, Satya Nagar, India}
\address[add2]{Indian Institute of Technology Madras, Chennai 600036, India}
\address[add3]{Indian Institute of Technology Guwahati, Assam 781039, India}
\address[add4]{Kavli Institute for the Physics and Mathematics of the Universe (WPI), University of Tokyo, Kashiwa 277-8583, Japan}
\address[add5]{High Energy Accelerator Research Organization (KEK), Tsukuba 305-0801, Japan, $^{A}$deceased}
\address[add6]{H. Niewodniczanski Institute of Nuclear Physics, Krakow 31-342, Poland}
\address[add7]{Department of Physics, Kyungpook National University, Daegu 702-701, Korea}
\address[add8]{School of Physics, University of Melbourne, Melbourne, Victoria 3010, Australia}
\address[add9]{Department of Physics, Niigata University, Niigata 950-2181, Japan, $^B$presently at Kitasato University, Sagamihara 252-0373, Japan}
\address[add11]{Dipartimento di Fisica, Universit\`{a} di Pisa, I-56127 Pisa, Italy}
\address[add12]{INFN Sezione di Pisa, I-56127 Pisa, Italy}
\address[add13]{Faculty of Mathematics and Physics, Charles University, 121 16 Prague, Czech Republic}
\address[add14]{Department of Physics and Astronomy, Seoul National University, Seoul 151-742, Korea}
\address[add15]{Tata Institute of Fundamental Research, Mumbai 400005, India, $^C$also at Utkal University, Bhubaneswar 751004, India}
\address[add16]{Department of Physics, Tohoku University, Sendai 980-8578, Japan}
\address[add17]{Department of Physics, University of Tokyo, Tokyo 113-0033, Japan}
\address[add18]{Dipartimento di Fisica, Universit\`{a} di Trieste, I-34127 Trieste, Italy}
\address[add19]{INFN Sezione di Trieste, I-34127 Trieste, Italy, $^D$presently at TIFPA - INFN, I-38123 Trento, Italy}
\address[add20]{Institute of High Energy Physics, Austrian Academy of Sciences, 1050 Vienna, Austria}

\begin{abstract}
The Belle\,II experiment at the SuperKEKB collider in Japan is designed to indirectly probe new physics using approximately
$50$ times the data recorded by its predecessor.
An accurate determination of the decay-point position of subatomic particles such as beauty and charm hadrons as well as a
precise measurement of low-momentum charged particles will play a key role in this pursuit.
These will be accomplished by an inner tracking device comprising two layers of pixelated silicon detector and four layers of
silicon vertex detector based on double-sided microstrip sensors.
We describe herein the design, prototyping and construction efforts of the Belle-II silicon vertex detector.
\end{abstract}


\end{frontmatter}

\section{Introduction}
\label{sec1}

The discovery of a new particle, which is consistent with the Higgs boson within the current experimental uncertainties,
by the ATLAS~\cite{higgs1} and CMS~\cite{higgs2} experiments at the LHC seems to complete the story of the standard
model. The focus is now geared towards deciphering the next fundamental layer of physics, often referred to as ``new
physics'', as the standard model has several shortcomings. For instance, it does not have a suitable dark-matter candidate,
nor can it fully account for the observed matter-antimatter asymmetry in universe.

As a next-generation flavor experiment, Belle\,II~\cite{belle2tdr} at the SuperKEKB collider~\cite{superkekb} is expected
to play a pivotal role in the above pursuit. Using a huge sample of $e^+e^-$ collision data recorded at various $\YnS$
resonances, about $50$ times that of its predecessor experiment (Belle~\cite{belleptp}), it will indirectly probe new physics
at an unprecedented level.
Measurements of charge-parity violation asymmetry in the decays of beauty and charm hadrons constitute a key experimental
approach for Belle\,II. The studies hinge on an accurate determination of decay-point positions of these hadrons as well as
a precise measurement of low-momentum charged particles.
These important tasks are accomplished by a sophisticated inner tracking device comprising two layers of pixelated silicon
detector (PXD) and four layers of silicon vertex detector (SVD) based on double-sided microstrip sensors.

We report herein the design, prototyping and construction efforts of the Belle-II SVD. Details on the PXD can be found
at Ref.~\cite{pxd-hsd2015}.

\section{Silicon Vertex Detector}
\label{sec2}

The PXD and SVD are the two innermost subdetectors of the Belle\,II experiment.
They nicely complement each other in providing information of excellent spatial granularity and timing resolution.
The resulting impact parameter resolution improves by a factor of two compared to Belle, from 40 down to 20\,$\mu$m~\cite{belle2tdr}.
Furthermore, the reconstruction efficiency of low-momentum particles and relatively long-lived particles such as $\KS$
mesons is improved, thanks to the extended outermost SVD layer.
The radii of the inner two PXD layers (Layer 1 and 2) are $14$ and $22$\,mm, while the radii of the outer four SVD
layers (Layer 3 through 6) are $38$, $80$, $115$, and $140$\,mm.
In comparison, Belle had the outermost SVD layer at a radius of $88$\,mm~\cite{bellesvd2}.

\begin{figure}[!htb]
\centering
\includegraphics[width=0.99\linewidth]{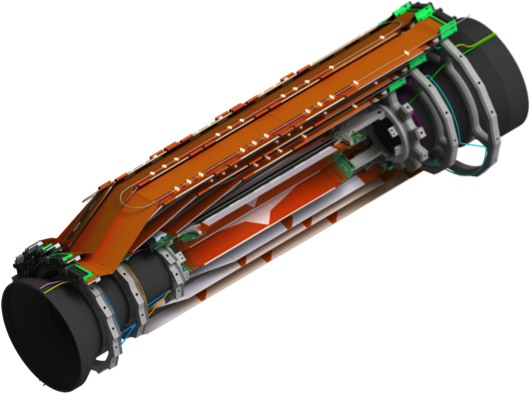}
\caption{A cut-off view of the Belle-II silicon vertex detector.}
\label{fig:layout}
\end{figure}

Figure~\ref{fig:layout} shows a cut-off view of the SVD. Its L3, L4, L5, and L6 (L stands for Layer) are respectively
composed of $7$, $10$, $12$, and $14$ modules or `ladders' with a polar-angle coverage of $17^\circ<\theta<
150^\circ$.
The asymmetry reflects the forward boost of the center-of-mass system arising due to a difference in energies
($4$ on $7$\,GeV) of the colliding $e^+$ and $e^-$ beams.
The L3 ladders are straight in shape, while those for L4-6 exhibit a slanted structure in the forward side.
The corresponding slant angles are $11.9^\circ$, $17.2^\circ$ and $21.1^\circ$.
The idea behind this lantern-like layout is to reduce the material budget and number of readout channels without compromising
the performance.
Clearly, these benefits come at the cost of a fairly complex mechanical structure.

\section{Sensors and Readout ASIC}
\label{sec3}

Three types of double-sided silicon microstrip detectors (DSSDs) are used in the SVD, all having the same length
of 12.3\,cm and thickness of 300 or 320\,$\mu$m.
The large rectangular DSSDs are $5.8$\,cm wide and fully utilize the available space on a 6-inch wafer.
A smaller width ($3.8$\,cm) is chosen for L3 to ensure that the DSSDs are aranged in a circular fashion around the PXD.
The width of the trapezoidal DSSDs in the forward part ranges between $3.8$ to $5.8$\,cm.
Table~\ref{tab:sensor} lists the specifications of various DSSDs used in the SVD.

\begin{table}[tbh]
\caption{Specifications of various DSSDs used in the SVD.}
\label{tab:sensor}
\resizebox{\columnwidth}{!}{
\begin{tabular}{cccccc}
\hline\hline
Type & \# strip & \# strip & Strip pitch & Strip pitch & Active area \\
     & p-side & n-side &  p-side ($\mu$m) & n-side ($\mu$m) & (${\rm mm}^2$)  \\
\hline
Small & $768$ & $768$ & $50$ & $160$ & $4716$ \\
Large & $768$ & $512$ & $75$ & $240$ & $7073$ \\
Trape. & $768$ & $512$ & $50$-$75$ & $240$ & $5890$\\
\hline
\end{tabular}}
\end{table}

The longer strips on the p-side are placed along the $z$ axis (parallel to the beam direction) and the shorter
strips on the n-side are located on the transverse $x$-$y$ plane.
The p-side of the L4-6 DSSDs faces the beam pipe, while the L3 DSSDs are oppositely arranged.
Both small and large rectangular DSSDs are fabricated by Hamamatsu Photonics in Japan, and the trapezoidal
ones by Micron Semiconductor in the UK.

To cope with the high particle rates expected at Belle\,II, the DSSDs require fast readout electronics with a short
integration (shaping) time.
A pipeline is also necessary to allow a dead-time free data taking.
Furthermore, a radiation tolerance up to $100$\,kGy is desirable.
The APV25 readout chip~\cite{apv25}, originally developed for CMS, is found to satisfy all these requirements.
It has a shaping time of $50$\,ns, compared to $800$\,ns of the VA1TA chip used in Belle, a pipeline depth
of $192$ cells, and a nominal clock speed of $40$\,MHz.
These chips are fabricated using a $0.25\,\mu$m CMOS process and can withstand a radiation dose of over
$1$\,MGy.

Due to faster shaping of the APV25 chip compared to VA1TA,
a smaller input capacitance is required for the amplifier to retain signal-to-noise ratio.
This also means that concatenations of several DSSDs, as was done for Belle, are simply prohibitive necessitating
the need to place the readout ASIC as close to the DSSD as possible.
These conflicting requirements are met by the so-called ``origami'' chip-on-sensor design~\cite{origami}.

\section{Origami Chip-on-sensor Design}
\label{sec4}

The full L3 ladder, and the forward and backward DSSDs in L4-6 adopt an edge-side readout scheme
with fanouts, similar to Belle.
On the other hand, the central DSSDs (one, two, and three for an L4, L5, and L6 ladder, respectively) rely on the
origami concept.
This novel scheme helps minimize the distance between the DSSD and the readout electronics, which in turn
reduces the capacitive noise.
The origami is a three-layer flexible fanout circuit (FlexPA) made of polymide.
It has ten APV25 chips for reading out the sensors.
Table~\ref{tab:ladder} describes the number of ladders for each SVD layer, the number of DSSDs and origamis
per ladder, and the number of APV25 chips per DSSD. 

\begin{table}[tbh]
\caption{Number of ladders, DSSDs, origamis and APV25 chips required for different SVD layers.}
\label{tab:ladder}
\resizebox{\columnwidth}{!}{
\begin{tabular}{ccccc}
\hline\hline
Layer & Ladders & DSSDs/ladder & Origamis/ladder & APVs/DSSD \\
\hline
L3 & 7 & 2 & 0 & 12\\
L4 & 10 & 3 & 1 & 10\\
L5 & 12 & 4 & 2 & 10\\
L6 & 16 & 5 & 3 & 10\\
\hline
\end{tabular}}
\end{table}

\begin{figure}[!htb]
\centering
\includegraphics[width=0.99\linewidth]{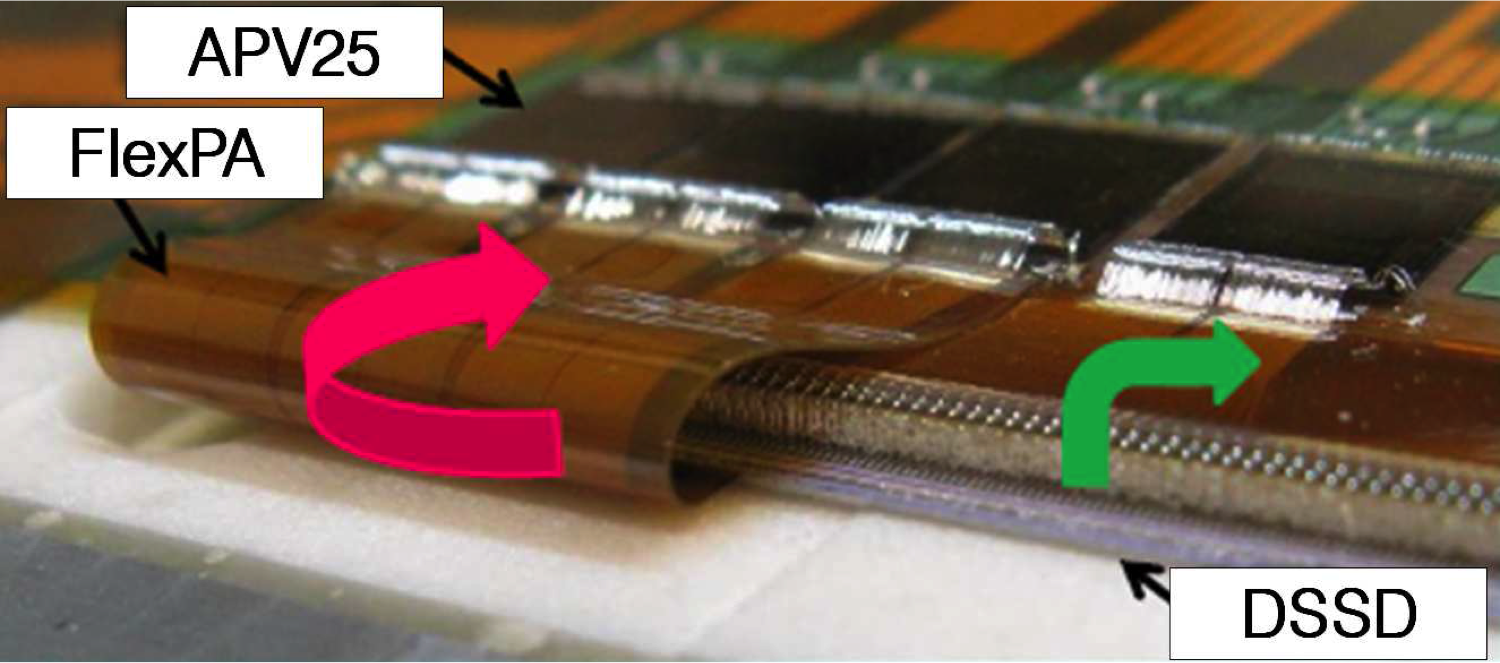}
\caption{The bent flex circuit transmitting the p-side signals (bottom) to the APV25 chips, which are glued onto the
origami FlexPA on the n-side (top). The green arrow shows the n-side wire bonding connections to the chips.}
\label{fig:origami}
\end{figure}

We need three different kinds of origami FlexPA designs depending on the SVD layer: $-z$ (backward), ce (central),
and $+z$ (forward).
All three of them are electrically equivalent differing only in the length of the tail part, or in the outer shape in case of
origami $+z$, which is only used in L6.
Between the FlexPA and DSSDs, a $1$\,mm thick sheet of light-weight styrofoam (Airex~\cite{airex}) is inserted.
This sheet provides both thermal and electrical insulation to minimize the heat transfer from the chips to the DSSDs
and to avoid signal cross-talk.
The APV25 chips are thinned down to 100\,$\mu$m for the material budget reduction, and are glued to the origami
FlexPA by a thin layer of conductive adhesive.
The n-side strips of a DSSD are directly connected to the chips by wire-bonding and a small fanout circuit.
Those of the p-side are attached by two flexible fanout circuits, which are bent around the edge of the DSSD and
glued onto the origami FlexPA in front of the APV25 chips.
A gluing robot is used to dispense the glue, while an ultrasonic wedge bonder used for wire-bonding.
Figure~\ref{fig:origami} shows the key feature of the origami concept.

\section{Assembly Procedure}
\label{sec5}

The ladder assembly procedure~\cite{assembly} is complicated and requires several kinds of jigs for each
operation involved.
The main purpose of these jigs is to ensure a precise alignment of the DSSDs during assembly.
The total number of jigs required is different for each layer, being the maximum for L6.

For the ladder assembly, we align the central DSSD(s) with the p-side at top, glue two fanout circuits on it,
and perform wire-bonding between the readout and fanout circuit pads.
We then flip the DSSD and place it on the assembly bench.
At this stage, we pickup the forward and backward subassemblies from the multipurpose chuck
and place them on the assembly bench.
We use a precision three-dimensional coordinate measuring machine (CMM) to align the sensors
within a tolerance of 10\,$\mu$m.
Each ladder is supported by two ribs made of carbon-fiber reinforced Airex foam.
The rib structure is very light but extremely stiff.
We pick up the forward and backward subassemblies from the assembly bench, and glue them on the ribs.
While assembling the central DSSD we first glue the Airex on the sensor, glue the origami module on the Airex,
and then perform wirebonding.
The pick-up and placing of the sensors and flex circuits are accomplished via vacuum chucking.
At the end, we place the plastic clips for holding the cooling pipe (see below).
Figure~\ref{fig:assembly} illustrates a part of the ladder assembly procedure.

\begin{figure}[!htb]
\centering
\includegraphics[width=0.39\linewidth]{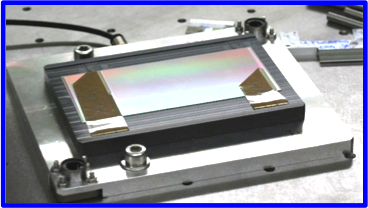}
\includegraphics[width=0.285\linewidth]{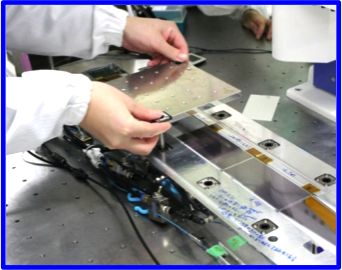}
\includegraphics[width=0.29\linewidth]{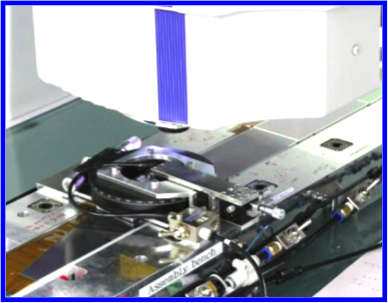}
\caption{A part of the ladder assembly procedure: (left) a DSSD with flex circuits vacuum chucked on the sensor
jig, (middle) the DSSD with flex circuits placed on the assembly bench, and (right) the DSSD aligned with the
XYZ$\theta$ stage under the CMM.}
\label{fig:assembly}
\end{figure}

On top of all APVs, we place a sheet of glass fiber and silicone rubber composite
(Keratherm~\cite{keratherm}) that is electrically isolated but thermally conductive.
The SVD power dissipation per active area is estimated to be 60.7\,mW/cm$^2$. 
The APV25 chips are cooled down to $-20^\circ$\,C with two-phase CO$_2$ flowing inside
a thin $1.6$-mm diameter pipe that is placed on Keratherm.
This cooling has two major advantages: a) it can withstand large amount of radiation
and b) it exhibits an excellent thermomechanical behavior.
Both the PXD and SVD system share a common volume, filled with dry nitrogen gas
to maintain a stable temperature and low dew point for avoiding condensation.
The average material budget for one ladder including ribs, DSSDs, electronics and cooling
is about $0.7\%$ of a radiation length.

\begin{figure}[!htb]
\centering
\includegraphics[width=0.99\linewidth]{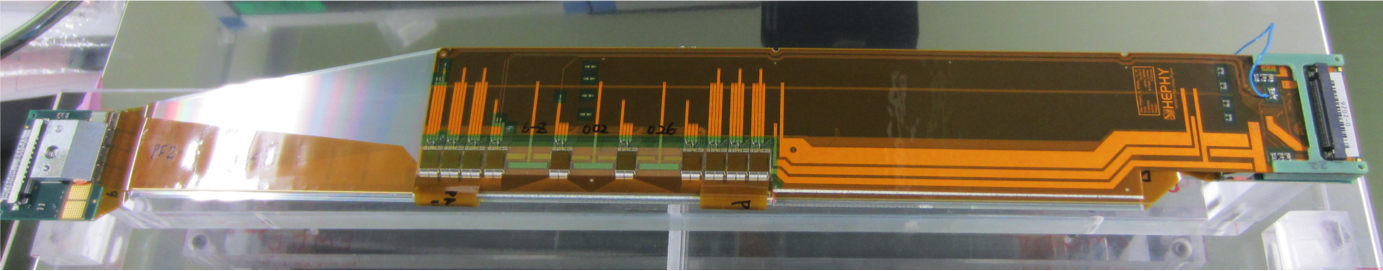}
\caption{An L4 mechanical prototype ladder.}
\label{fig:l4ladder}
\end{figure}

\begin{figure}[!htb]
\centering
\includegraphics[width=0.41\linewidth]{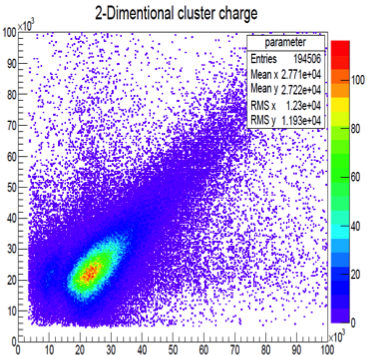}
\includegraphics[width=0.58\linewidth]{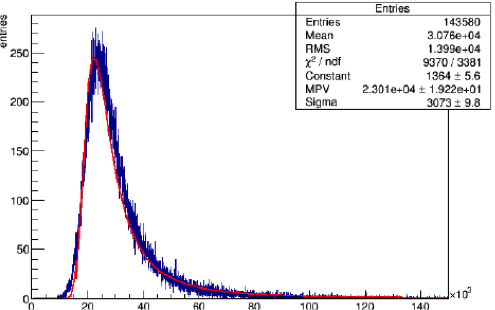}
\caption{Results of the source measurement performed on an electrically working L5 ladder: (left) the correlation
of cluster charge between the p- and n-side, and (right) the cluster charge on the p-side fitted with a Landau distribution.}
\label{fig:eqa}
\end{figure}

The construction workflow is split among several sites.
The forward and backward subassemblies are produced at INFN Pisa
and later shipped to other assembly sites: L4 (TIFR), L5 (HEPHY), and L6 (IPMU).
The L3 assembly is performed by the Melbourne group.
The R\&D for the ladder assembly procedure is now over, and each site has
assembled a number of mechanical prototypes as well as one or two electrically working ladders.
The latter have been tested with a $^{90}\mbox{Sr}$ $\beta$-ray source or laser to assess the overall
performance and potential defects of the sensors.
We present an L4 mechanical prototype in Figure~\ref{fig:l4ladder}. Results of the source measurement
performed on an electrically working L5 ladder are shown in Figure~\ref{fig:eqa}.

\section{Backend Electronics}
\label{sec6}

\begin{figure}[!htb]
\centering
\includegraphics[width=0.99\linewidth]{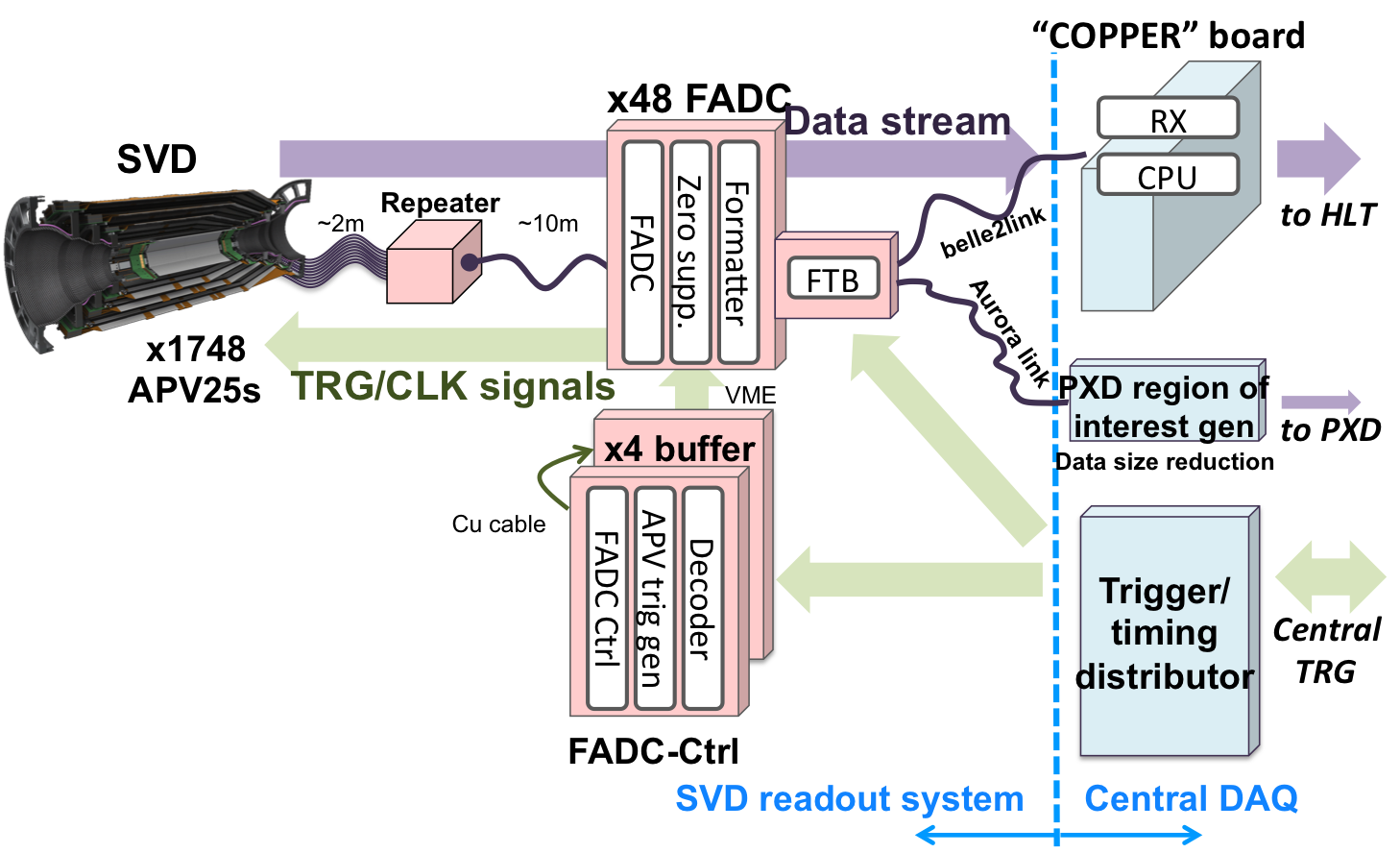}
\caption{A schematic view of the SVD readout system.}
\label{fig:electronics}
\end{figure}

Figure~\ref{fig:electronics} shows an overview of the whole SVD readout system~\cite{readout}.
The analog signals from the APV25 chips are transmitted to flash analog-to-digital converter (FADC) boards
via repeater boards.
Each FADC board can receive up to 48 APV25 analog outputs, and performs the analog to digital conversion for
obtaining digitized signals.
The digitized signals are decoded and further processed on an FPGA, and propagated to a finesse transmitter
board (FTB).
The FTB sends the data to the common pipelined platform for electronics readout (COPPER), which is the
Belle\,II DAQ interface, through an optical cable using a unified high-speed serial protocol (belle2link).
In parallel, the FTB also transmits a replica output to the PXD data acquistion system using an Aurora link.
The COPPER board performs further data processing and sends the data for storage via the high-level
trigger (HLT) system.

A total of 48 FADC boards and 48 FTBs will be employed across Belle\,II.
All the components in the SVD readout system are being developed while prototypes were successfully
produced and tested.

\section{Beam Test Results}
\label{sec7}

A test of the complete PXD+SVD readout chain was performed~\cite{beamtest} using the electron beam at DESY.
The beam energy was 2 to 6\,GeV and 1\,T of magnetic field was applied perpendicular to the beam line.
The set-up included four SVD test modules, with one large rectangular DSSD in each, and one PXD
module in a light-tight box, FADC and FTB boards, CO$_2$ cooling, slow control and environmental
sensors based on optical fiber sensors.
Several aspects were checked.
We find the SVD cluster hit efficiency to be above $99\%$ for tracks within the fiducial volume.
Further, we confirm that the common mode correction and zero suppression schemes do not deteriorate
the SVD hit efficiency.

\section{Conclusions}
\label{sec8}

In summary, the KEKB machine and Belle experiment are being upgraded to SuperKEKB and Belle\,II
with a goal to indirectly probe new physics, in which the SVD will play a key role.
We have developed a robust assembly procedure for the SVD ladder modules after extensive R\&D.
This includes the production of dedicated jigs, their fine-tuning and handling, glue spread control, and
wire-bonding parameter tuning.
The most innovative aspect among all has been the origami chip-on-sensor scheme that enables
an excellent signal-to-noise ratio by reducing the capacitive noise. The assembly procedure is found to
be reproducible allowing us to consistently assemble the ladders with an acceptable mechanical offset
and good electrical quality.

The full-scale production of SVD ladders will start at the beginning of 2016, which will be followed by
commissioning in mid-2017.
The first physics run with both PXD and SVD in will be towards the end of 2018.

\section*{Acknowledgements}
We congratulate the organizers for a well organized conference and are grateful to colleagues those
have helped us in preparing these proceedings. The research leading to these results has received funding
from the European Commission under the FP7 Research Infrastructures project AIDA, grant agreement
no. 262025.

\end{document}